\documentclass[conference]{IEEEtran}
\usepackage{times}

\usepackage[numbers]{natbib}
\usepackage{multicol}
\usepackage[bookmarks=true]{hyperref}

\usepackage{subcaption}
\usepackage{float}
\usepackage{amsmath}
\usepackage{amssymb}
\usepackage{xcolor}
\usepackage{graphicx}
\usepackage[capitalize]{cleveref}

\begin{document}

\title{
Towards Zero-Shot Coordination between Teams of Agents: The N-XPlay Framework
 }

\author{
Ava Abderezaei*, Chi-Hui Lin*, Joseph Miceli, Naren Sivagnanadasan, \\
Stéphane Aroca-Ouellette, Jake Brawer, Alessandro Roncone
}

\maketitle

\begin{abstract}
Zero-shot coordination (ZSC)—the ability to collaborate with unfamiliar partners—is essential to making autonomous agents effective teammates. Existing ZSC methods evaluate coordination capabilities between two agents who have not previously interacted. However, these scenarios do not reflect the complexity of real-world multi-agent systems, where coordination often involves a hierarchy of sub-groups and interactions between teams of agents, known as Multi-Team Systems (MTS). To address this gap, we first introduce N-player Overcooked, an N-agent extension of the popular two-agent ZSC benchmark, enabling evaluation of ZSC in N-agent scenarios. We then propose N-XPlay for ZSC in N-agent, multi-team settings. Comparison against Self-Play across two-, three- and five-player Overcooked scenarios, where agents are split between an ``ego-team'' and a group of unseen collaborators shows that agents trained with N-XPlay are better able to simultaneously balance ``intra-team'' and ``inter-team'' coordination than agents trained with SP.
\end{abstract}

\IEEEpeerreviewmaketitle

\section{Introduction}
Zero-Shot Coordination (ZSC) is an open-problem in multi-agent systems, which challenges agents to efficiently and robustly collaborate with previously unseen teammates \cite{peterstone}. Existing work on ZSC mostly centers on systems of two agents. One agent---often referred to as the ego-agent---will be trained and evaluated on its ability to coordinate with unseen partners, such as a human collaborator, in two-agent teams  \cite{gamma,pecan,fcp}. However, many real-world scenarios involve coordination of many agents across a hierarchy of sub-groups \cite{lin2024}. Agents may have advanced capabilities within particular local teams, yet they must be able to work with members of other groups to accomplish organization-level goals. Humans manage these different relationships---namely, intra-team and inter-team---in multi-team systems (MTS) \cite{teamsofteams3}, and are able to decompose tasks within a sub-group and across the hierarchy of sub-groups. This MTS formulation of the ZSC problem---where agents must balance coordination within an `ego-team' while still effectively collaborating with a group of previously unseen teammates---is underexplored. The ability to accommodate multiple forms of coordination or generally multiple strategies simultaneously has implications beyond MTS, and will be a necessary feature of agents deployed in large scale, heterogeneous organizations such as Human-AI teams. 

\begin{figure}
    \centering
    \includegraphics[width=0.75\linewidth]{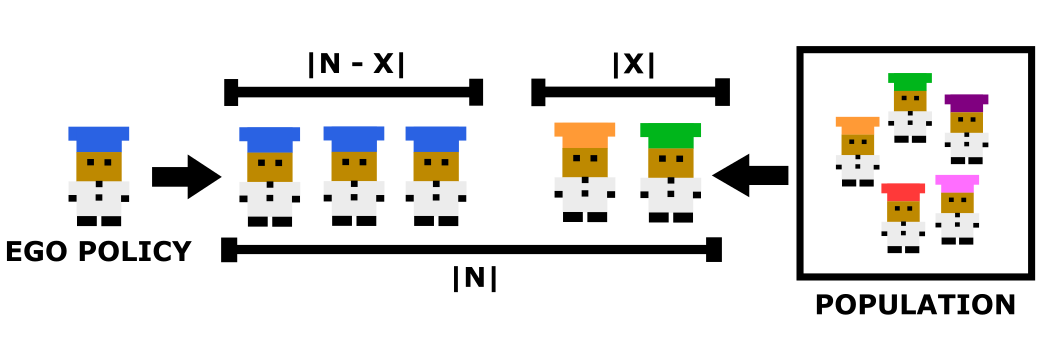}
    \caption{\textbf{N-XPlay}: The ego-team (blue) consists of $N-X$ agents using identical policies, aiming to maximize intra-team performance while coordinating with $X$ collaborators. During training, collaborators are sampled from a population-based method such as \cite{fcp,pecan, trajedi} to enable inter-team collaboration.}
    \label{fig:n-2}
    \vspace{-15pt}
\end{figure}

To lay the groundwork for studying ZSC in more complex organizational structures, we extend the the popular ZSC environment Overcooked \cite{carroll2020utilitylearninghumanshumanai} from a two-player game to an N-player game. Overcooked is a timed collaborative game where agents must coordinate to prepare and deliver meals. In expanding Overcooked to N-players, complexities not seen in dyadic systems begin to appear such as the structure and composition of teams. These complexities exacerbate the ZSC problem, creating a gap between current techniques and their applicability to settings with more than two agents. To address this gap, we propose \textsl{N-XPlay}, a framework to train agents for use in multi-team systems. This approach is motivated by the ``Teams of Teams'' (\cite{teamsofteams1}) concept, where disparate groups must operate effectively both within and across teams. N-XPlay (\cref{fig:n-2}) naturally extends the existing body of work in ZSC from two-agent to N-agent settings to learn a policy that can both collaborate with replicas of itself and unseen collaborators simultaneously. In training, a subset of $N - X$ agents sharing a policy (similar to Self-Play; SP) are paired with $X$ agents independently sampled from a diverse population of pre-trained policies, similar to existing Population-Based (PB) methods \cite{pecan, gamma, fcp}. This combination of SP and PB enables learning both intra-team coordination within the ``ego-team'' and inter-team coordination with an unseen group of agents.

We analyze N-XPlay on our N-player Overcooked environment in two-, three-, and five-player settings. We compare the performance of ego-teams comprising of agents trained using N-XPlay against ego-teams composed of agents trained using N-agent SP when paired with varying numbers of unseen collaborators. In summary, our contributions are twofold: (1) we extend the Overcooked environment popularly used for two-agent ZSC tasks, to the N-agent setting \footnote{Open-source code available at: \href{https://github.com/HIRO-group/multiHRI}{https://github.com/HIRO-group/multiHRI}.} and, (2) we propose N-XPlay, a method that extends existing two-agent ZSC approaches to N-agent multi-team settings.
\begin{figure}
    \centering
    \includegraphics[width=\linewidth]{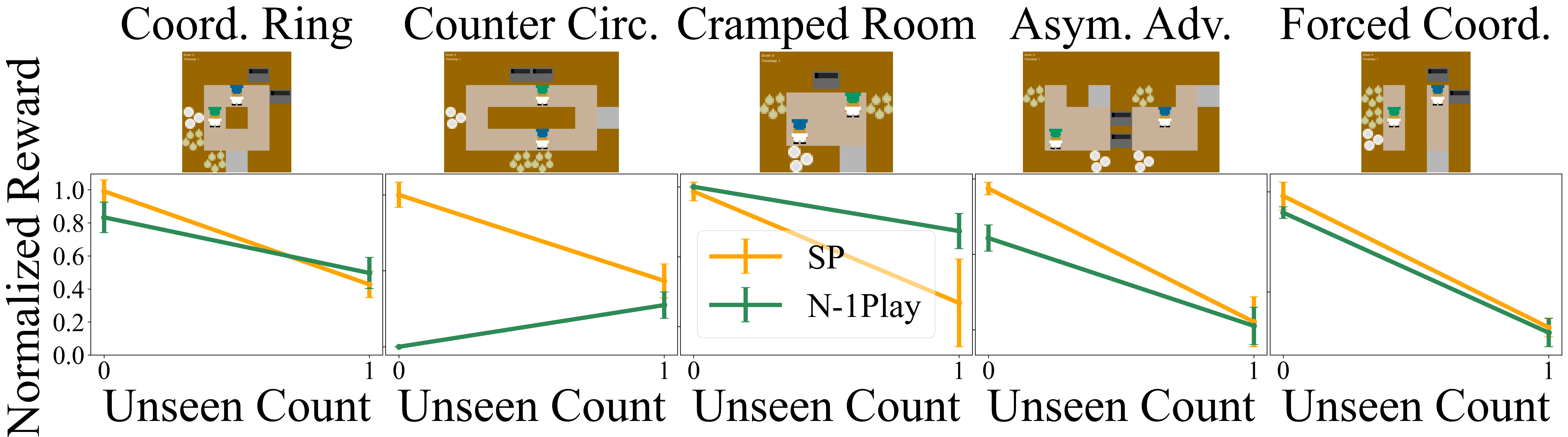}
    \caption{\textbf{N-1Play vs SP on two-agent layouts:} With no unseen teammate, SP outperforms N-1Play. With 1 unseen, SP degrades, and N-1Play outperforms it on two layouts.}
    \label{fig:2-results}
    \vspace{-10pt}
\end{figure}
\section{Method and Experiments}
\textbf{\textsl{N-XPlay Framework.}}
N-XPlay extends two-agent ZSC methods to N-agent settings, addressing challenges presented in complex organizational structures. It models these organization compositions using a ``Teams of Teams'' (\cite{teamsofteams1}) approach, creating agents that coordinate effectively with each other while capable of collaborating with unknown agents. To do so, it first creates a population of diverse reinforcement learning agents using existing population-based methods such as FCP \cite{fcp} and MEP \cite{zhao2021maximum}. In training the ego-team policy for an $N$-agent setting, for each episode, a subset of agents sized $N - X$ share the same policy similar to SP, while paired with $X$ agents independently sampled from the population. As an example in \cref{fig:n-2}, in a environment of $N = 5$ agents where $X = 2$, during training 3 teammates will share and learn the same policy and 2 will be sampled from the pre-trained population.

\begin{figure}
    \centering
    \includegraphics[width=0.75\linewidth]{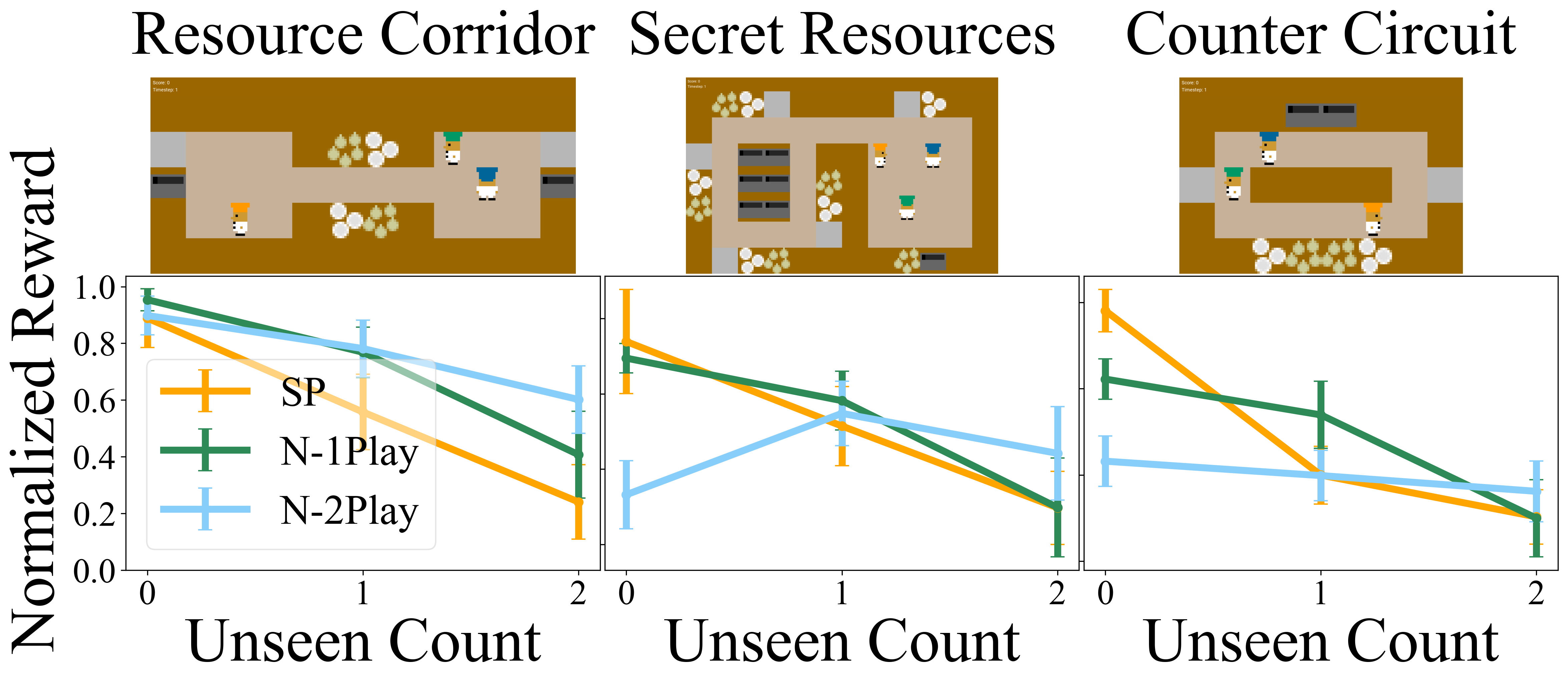}
    \caption{\textbf{N-XPlay vs SP on three-agent layouts}: As the number of unseen teammates increases, performances shift: N-1Play is best with one unseen teammate, N-2Play is best with two.}
    \label{fig:3-results}
\end{figure}

\textbf{\textsl{Experiments.}}
With our experiments, we wished to explore the impact of the N-XPlay training regime on team performance across various team compositions and sizes. To do this, we build on top of \cite{stephane, stephane2}'s implementation and generate an unseen collaborator population similar to FCP \cite{fcp} where we train four SP agents and store three checkpoints during their training and categorize them into---high, medium, and low performance---based on their average rewards. N-XPlay policies are then trained using the method described above. We then evaluate collective performance in Overcooked where some proportion of the agents are N-XPlay or agents trained with N-player SP and the remaining agents are drawn from a previously unseen population.
For a thorough analysis, we evaluate N-XPlay across various organization sizes and team compositions:
(i) Two-player: N-1Play, with one agent trained alongside a sampled teammate. 
(ii) Three-player: N-1Play (two trained, one sampled) and N-2Play (one trained, two sampled).
(iii) Five-player: N-1, N-3, and N-4Play, training four, two, and one agent(s), with the rest sampled from the population. \cref{fig:2-results}, \cref{fig:3-results}, and \cref{fig:5-results} compare the performance of SP and versions of N-XPlay across these layouts when paired with varying numbers of unseen teammates.
\section{Discussion and Conclusion}
\label{sec:conclusion}
\begin{figure}
    \centering
    \includegraphics[width=\linewidth]{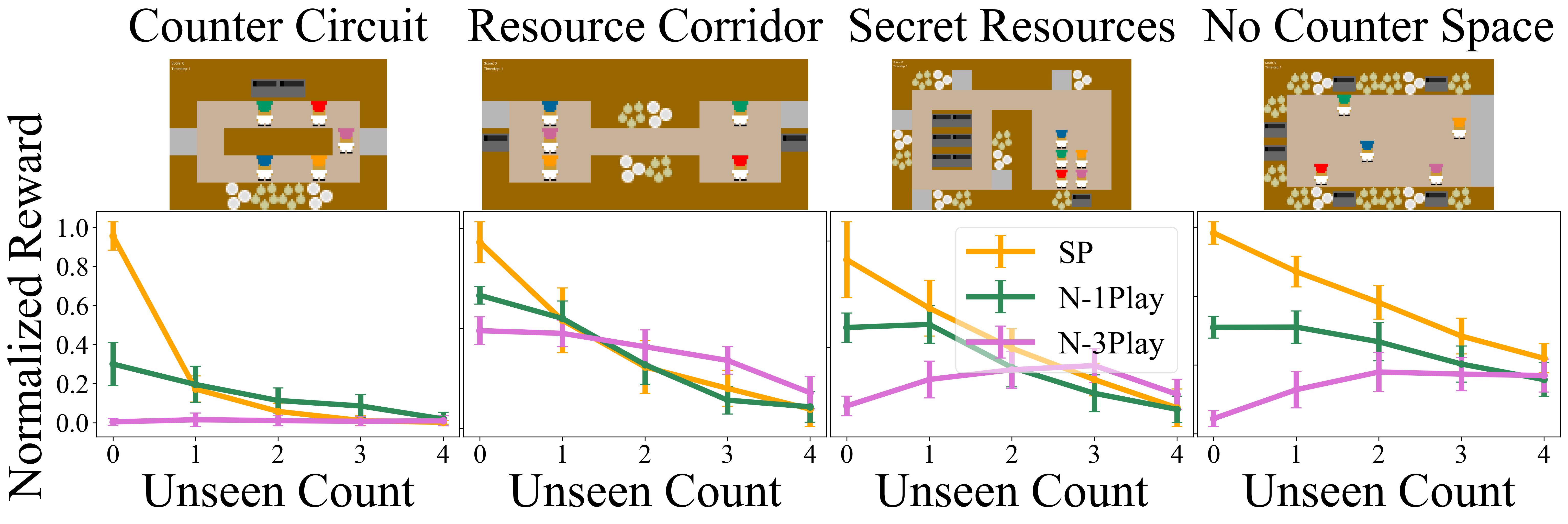}
    \caption{\textbf{N-XPlay vs SP on five-agent layouts:} Layout configuration significantly impacts the evaluation. N-XPlay lags behind SP in low-coordination layouts like No Counter Space but outperforms SP in high-coordination settings.}
    \label{fig:5-results}
    \vspace{-15pt}
\end{figure}

\begin{figure}
    \vspace{-15pt}
    \centering
\includegraphics[width=0.8\linewidth]{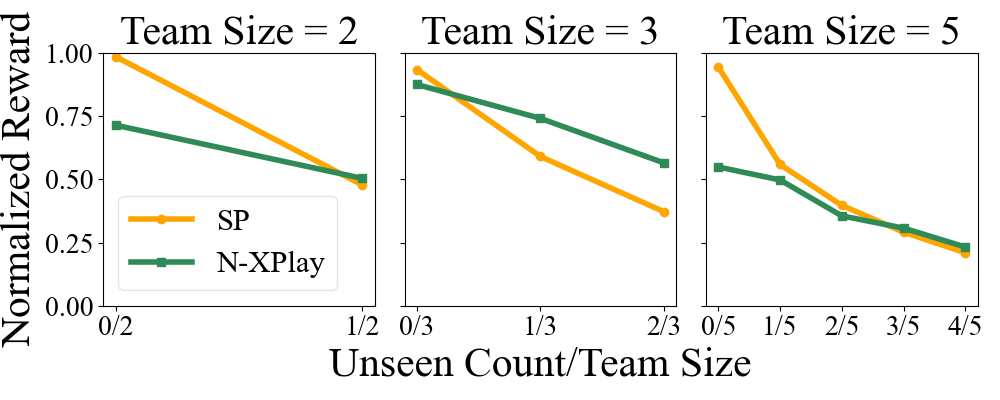}
    \caption{
    \textbf{Best performing variant of N-XPlay vs SP:} As the ratio of unseen teammates (X/N) increases, SP and N-XPlay performance decline with N-XPlay ultimately surpassing SP.}
    \label{fig:ratio-unseen-over-teamsize}
    \vspace{-15pt}
\end{figure}

To examine the effect of organization composition, \cref{fig:ratio-unseen-over-teamsize} plots the performance of SP and the best-performing variant of N-XPlay (referred to as N-XPlay in the rest of this section), averaged across layouts, against the unseen-to-total agent ratio. As this ratio increases i.e., less of the organization is represented by the ego-team, N-XPlay begins to outperform SP. For instance, N-XPlay outperforms SP with one unseen teammate in two-agent teams ($1/2$), but not in five-agent teams ($1/5$). In the latter, N-XPlay only overtakes SP when at least three teammates are unseen ($3/5$). We also see some layouts, like No Counter Space (\cref{fig:5-results}), require little coordination and allow agents to work independently, allowing SP to excel regardless of unseen teammates.
Our results reveal two key conclusions to be expanded on in future work: (1) As the ratio of unseen over collective size increases (i.e. the ego-team represents a smaller portion of the collective), the performance of both SP and N-XPlay declines with N-XPlay starting to outperform SP in layouts requiring coordination among the collective. (2) Focusing solely on maximizing intra-team performance, as in the case of SP, increases the collective’s dependence on the ego-team's ability to independently perform the org-level task. When the ego-team cannot accomplish the overall task alone, the absence of inter-team coordination leads to substantial performance loss. Therefore, the ability to coordinate both within and across teams is critical in complex MTS settings.

To conclude, here we introduced N-player Overcooked---an extension of the widely used two-player Overcooked environment---to enable the study of zero-shot coordination (ZSC) in settings with more than two agents. Building on this, we proposed N-XPlay, a framework that generalizes ZSC to N-agent settings by modeling the complex team compositions that arise in larger groups. Our results demonstrate N-XPlay is better suited than SP at finding policies that can simultaneously perform both intra- and inter-team collaboration.

\bibliographystyle{plainnat}
\bibliography{references}

\begin{thebibliography}{12}
\providecommand{\natexlab}[1]{#1}
\providecommand{\url}[1]{\texttt{#1}}
\expandafter\ifx\csname urlstyle\endcsname\relax
  \providecommand{\doi}[1]{doi: #1}\else
  \providecommand{\doi}{doi: \begingroup \urlstyle{rm}\Url}\fi

\bibitem[Aroca-Ouellette et~al.(2023)Aroca-Ouellette, Aroca-Ouellette, Biswas, Kann, and Roncone]{stephane}
St\'{e}phane Aroca-Ouellette, Miguel Aroca-Ouellette, Upasana Biswas, Katharina Kann, and Alessandro Roncone.
\newblock Hierarchical reinforcement learning for ad hoc teaming.
\newblock In \emph{Proceedings of the 2023 International Conference on Autonomous Agents and Multiagent Systems}, AAMAS '23, page 2337–2339, Richland, SC, 2023. International Foundation for Autonomous Agents and Multiagent Systems.
\newblock ISBN 9781450394321.

\bibitem[Aroca-Ouellette et~al.(2025)Aroca-Ouellette, Aroca-Ouellette, von~der Wense, and Roncone]{stephane2}
St{\'e}phane Aroca-Ouellette, Miguel Aroca-Ouellette, Katharina von~der Wense, and Alessandro Roncone.
\newblock Implicitly aligning humans and autonomous agents through shared task abstractions.
\newblock In \emph{Proceedings of the Thirty-Fourth International Joint Conference on Artificial Intelligence, {IJCAI-25}}. International Joint Conferences on Artificial Intelligence Organization, 2025.
\newblock Main Track.

\bibitem[Carroll et~al.(2019)Carroll, Shah, Ho, Griffiths, Seshia, Abbeel, and Dragan]{carroll2020utilitylearninghumanshumanai}
Micah Carroll, Rohin Shah, Mark~K Ho, Tom Griffiths, Sanjit Seshia, Pieter Abbeel, and Anca Dragan.
\newblock On the utility of learning about humans for human-ai coordination.
\newblock \emph{Advances in neural information processing systems}, 32, 2019.

\bibitem[Liang et~al.(2024)Liang, Chen, Gupta, Du, and Jaques]{gamma}
Yancheng Liang, Daphne Chen, Abhishek Gupta, Simon~Shaolei Du, and Natasha Jaques.
\newblock Learning to cooperate with humans using generative agents.
\newblock In \emph{The Thirty-eighth Annual Conference on Neural Information Processing Systems}, 2024.

\bibitem[Lin et~al.(2024)Lin, Koh, Roncone, and Chen]{lin2024}
Chi-Hui Lin, Joewie~J. Koh, Alessandro Roncone, and Lijun Chen.
\newblock {ROMA-iQSS}: An objective alignment approach via state-based value learning and round-robin multi-agent scheduling.
\newblock In \emph{2024 American Control Conference (ACC)}, pages 1083--1090, 2024.
\newblock \doi{10.23919/ACC60939.2024.10644908}.

\bibitem[Lou et~al.(2023)Lou, Guo, Zhang, Wang, Huang, and Du]{pecan}
Xingzhou Lou, Jiaxian Guo, Junge Zhang, Jun Wang, Kaiqi Huang, and Yali Du.
\newblock Pecan: Leveraging policy ensemble for context-aware zero-shot human-ai coordination.
\newblock In \emph{Proceedings of the 2023 International Conference on Autonomous Agents and Multiagent Systems}, AAMAS '23, page 679–688, Richland, SC, 2023. International Foundation for Autonomous Agents and Multiagent Systems.
\newblock ISBN 9781450394321.

\bibitem[Lupu et~al.(2021)Lupu, Cui, Hu, and Foerster]{trajedi}
Andrei Lupu, Brandon Cui, Hengyuan Hu, and Jakob Foerster.
\newblock Trajectory diversity for zero-shot coordination.
\newblock In Marina Meila and Tong Zhang, editors, \emph{Proceedings of the 38th International Conference on Machine Learning}, volume 139 of \emph{Proceedings of Machine Learning Research}, pages 7204--7213. PMLR, 18--24 Jul 2021.

\bibitem[Salas et~al.(2008)Salas, Cooke, and Rosen]{teamsofteams1}
Eduardo Salas, Nancy Cooke, and Michael Rosen.
\newblock On teams, teamwork, and team performance: Discoveries and developments.
\newblock \emph{Human factors}, 50:\penalty0 540--7, 07 2008.

\bibitem[Stone et~al.(2010)Stone, Kaminka, Kraus, and Rosenschein]{peterstone}
Peter Stone, Gal~A. Kaminka, Sarit Kraus, and Jeffrey~S. Rosenschein.
\newblock Ad hoc autonomous agent teams: collaboration without pre-coordination.
\newblock In \emph{Proceedings of the Twenty-Fourth AAAI Conference on Artificial Intelligence}, AAAI'10, page 1504–1509. AAAI Press, 2010.

\bibitem[Strouse et~al.(2021)Strouse, McKee, Botvinick, Hughes, and Everett]{fcp}
DJ~Strouse, Kevin~R. McKee, Matt Botvinick, Edward Hughes, and Richard Everett.
\newblock Collaborating with humans without human data.
\newblock In \emph{Proceedings of the 35th International Conference on Neural Information Processing Systems}, NIPS '21, Red Hook, NY, USA, 2021. Curran Associates Inc.
\newblock ISBN 9781713845393.

\bibitem[Zaccaro et~al.(2020)Zaccaro, Dubrow, Torres, and Campbell]{teamsofteams3}
Stephen~J. Zaccaro, Samantha Dubrow, Elisa~M. Torres, and Lauren~N.P. Campbell.
\newblock Multiteam systems: An integrated review and comparison of different forms.
\newblock \emph{Annual Review of Organizational Psychology and Organizational Behavior}, 7\penalty0 (Volume 7, 2020):\penalty0 479--503, 2020.
\newblock ISSN 2327-0616.

\bibitem[Zhao et~al.(2023)Zhao, Song, Yuan, Hu, Gao, Wu, Sun, and Yang]{zhao2021maximum}
Rui Zhao, Jinming Song, Yufeng Yuan, Haifeng Hu, Yang Gao, Yi~Wu, Zhongqian Sun, and Wei Yang.
\newblock Maximum entropy population-based training for zero-shot human-ai coordination.
\newblock \emph{Proceedings of the AAAI Conference on Artificial Intelligence}, 37\penalty0 (5):\penalty0 6145--6153, Jun. 2023.

\end{thebibliography}

\end{document}